\documentstyle[12pt]{article}
\def\be{\begin{equation}}
\def\ee{\end{equation}}
        \title{New vacuum solutions
 of conformal Weyl gravity}
\author{V. Dzhunushaliev, H.-J. Schmidt}
\date{July 19, 1999}
\begin{document}
\maketitle
\centerline{Institut f\"ur Mathematik, Universit\"at
Potsdam \footnote{dzhun@rz.uni-potsdam.de \qquad permanent address:
Dept. Theor. Phys., Kyrgyz State National University, Bishkek 720024, 
Kyrgyzstan} }
\centerline{PF 601553, D-14415 Potsdam, Germany}
 \centerline{and}
\centerline{Institut f\"ur Theoretische Physik, Freie
Univers. Berlin \footnote{http://www.physik.fu-berlin.de/\~{}hjschmi \ \ 
 \quad  hjschmi@rz.uni-potsdam.de}}
\centerline{Arnimallee 14, D-14195 Berlin, Germany}
\begin{abstract}
The Bach equation, i.e., the 
 vacuum field equation following from the Lagrangian
$L=C_{ijkl}C^{ijkl}$, will be completely  solved for the
 case that the metric is conformally related to the 
cartesian product of two 2-spaces; this covers the  
spherically and the plane symmetric space-times as special
subcases.

\medskip

Contrary to other approaches, we make a covariant
2+2-decompo\-si\-tion
   of the field equation, and so we are able
to apply results from  2-dimensional gravity. Finally, some 
 cosmological solutions will be presented and discussed. 
 \end{abstract}
Preprint UNIPO-MATH-99-July-19, submitted to J. Math. Phys. 

PACS number(s): 04.50.+h, 98.80.Cq\\
 \vspace{5.mm} 
\noindent 
KEY:  Alternative Theories of
Gravity, Cosmology, Conformal invariance.

\section{Introduction}
\setcounter{equation}{0} 
We consider the Lagrangian 
\be
L \qquad = \qquad C_{ijkl}C^{ijkl}
\ee 
where $C^i_{\, \, jkl} $ is the conformally invariant 
Weyl tensor.  The variational derivative of $L \sqrt{-g}$
 (where $g$ is the determinant of the metric $g_{ij}$)
 gives rise to the Bach tensor [1] \footnote{more exactly:
$$B^{ij} = \frac{1}{\sqrt{-g}} \cdot 
\frac{\delta L \sqrt{-g}}{\delta g_{ij}}$$}
\be
B_{ij} \qquad = \qquad 2 \ C^{k \ \ l}_{\ ij \ ;lk}
\quad + \quad C^{k \ \ l}_{\ ij } \ R_{lk}
 \ee
The purpose of the present paper is to characterize  
several solutions of the Bach equation
 $B_{ij}=0$, and thereby we cover the 
spherically and the plane symmetric metrics. 
In other words: We look for vacuum solutions of
conformal Weyl gravity [2]. 

\section{Another form of the field equation}
Subtracting the divergence\footnote{which represents the 
Gauss-Bonnet term in 4 dimensions} 
\be
L_{GB} \quad =
 \quad R_{ijkl} \ R^{ijkl} \ - \  4 R_{ij} \ R^{ij} \ + \ R^2
\ee
from the Lagrangian $L$  eq. (1) we get
$$
\tilde L \qquad = \qquad 2 \ R_{ij} \ R^{ij} 
\quad - \quad \frac{2}{3} \ R^2
$$
The variation of $\tilde L \sqrt{-g}$ with respect to the
metric gives, of course, a vacuum equation identical to eq.
(2), but now in a form [3], where neither 
the Weyl tensor not the full Riemann tensor explicitly 
appear. 
\be
B_{ij}     \qquad = \qquad B^{(1)}_{ij}  \ + \ B^{(2)}_{ij}   
\ee
where
\be
   B^{(1)}_{ij}   \quad = \quad
- \Box R_{ij} + 2 R^{\ k}_{i \ ;jk} - \frac{2}{3} R_{;ij}
 + \frac{1}{6} g_{ij} \Box R
\ee
and
\be
   B^{(2)}_{ij}  \ = \ \frac{2}{3} R \ R_{ij}
- 2 R_{ik}R^k_j - \frac{1}{6} R^2 g_{ij}
+ \frac{1}{2} g_{ij} R_{kl} R^{kl} 
\ee
This form of the field equation is also given in [4]; the two
details where the equation given in [4] differs from ours are
explained as follows:  

1. Instead of $+ 2 R^{\ k}_{i \ ;jk}$ they write
$+  R^{\ k}_{i \ ;jk} +  R^{\ k}_{j \ ;ik}$.

\smallskip

\noindent
However, the tensor $ R^{\ k}_{i \ ;jk}$ is already
symmetric in $ij$ due to the Bianchi identity. 

2. In our eq. (4) the authors of ref. [4] write $-$ instead
of $+$. But this is only due to the different sign
conventions. Our conventions are defined by 
$R_{ij} = R^k_{\ ikj}$ and the condition that the Euclidean 
sphere has $R>0$.

\section{The trivial solutions}
If $R_{ij} = \lambda g_{ij}$ for any constant $\lambda$,
then by eqs. (4-6) we see that $B_{ij} = 0$ is identically 
fulfilled. In other words: Every Einstein space\footnote{i.e.
vacuum solutions of the Einstein field equation with arbitrary
value of $\Lambda$.} solves the Bach equation. From eqs. (1,2)
it becomes clear that the Bach equation is conformally 
invariant. So we get: If we apply a conformal
transformation to a solution, then the resulting space-time
solves the Bach equation, too. Combining both properties 
it proves useful to define:

\medskip

A solution of the Bach equation is called trivial if it is
conformally related to an Einstein space.\footnote{In
 [5], conditions have been found to decide whether a given 
space-time is conformally related to an Einstein space;
however, as already mentioned there, these conditions are
applicable only in such cases where a nonvanishing
scalar constructed from the Weyl tensor exists.} In [6], the spherically 
symmetric solutions of the Bach equation have been analyzed, and the 
result
 is: Every spherically symmetric solution of the Bach equation is 
almost everywhere trivial. The restriction ``almost everywhere''  refers 
to
 possibly existing hypersurfaces where the necessary conformal factor 
 becomes singular.  Further details about the Bach tensor can be 
found in [7]. 

\section{The 2$+$2 decomposition of the Bach equation} 

In this section we perform a 2$+$2 decomposition of the  metric,  and 
then 
we
apply results [8] from 2-dimensional gravity to solve the Bach 
equation. 

\subsection{The metric ansatz}
For the metric
\be
ds^2\quad = \quad g_{ij} \, dx^i \, dx^j \, , \qquad i,j=0, \dots 3
\ee
we make the following ansatz:
\be
ds^2\quad = \quad d\sigma^2 \ + \ d\tau^2
\ee
where $d\sigma^2$ and $d\tau^2$  are both 2-dimensional. The 
metric 
\be
d\sigma^2 \quad = \quad g_{AB} \, dx^A \, dx^B \, , \qquad A,B=0,1 
\ee
where
 $g_{AB}$ depends on the $x^A$ only, has curvature scalar $P$ 
and signature 
$(-+)$. The other 2-dimensional metric 
\be
d\tau^2 \quad =
 \quad g_{\alpha \beta}\, dx^{\alpha} \, dx^{\beta}\, , \qquad
 \alpha, \beta = 2,3
\ee
where $g_{\alpha \beta}$ depends on the $x^{\alpha}$ only, has 
curvature
scalar $Q$ and signature $(++)$. 

For the 4-dimensional metric (7) we get signature $(-+++)$ and 
curvature scalar 
$R$ via 
\be
R(x^i)\quad = \quad P(x^A) \ + \ Q(x^{\alpha}) .
\ee
\subsection{The Einstein spaces of this type}
The Einstein spaces of the type defined in section 4.1. are already 
known
for a long time,
 see [9] for the history of these metrics. Here we deduce them for 
two reasons: First, 
we want to elucidate the method which we will apply
to   the Bach equation afterwards, and second, it is not yet general
 knowledge, that  
a spherically symmetric Einstein space (eq. (16) below) exists
 which cannot be written  in Schwarzschild coordinates.
The Einstein spaces can be found as extremals of the action
\be
I\quad = \quad \int (R \ - \ 2 \, \Lambda) \sqrt{- \det g_{ij}}\, d^4 x^i
\ee
where $\Lambda$ has an arbitrary constant value. Extremality implies 
constancy of
$R$, and because of eq. (11), we find both $P$ and $Q$ as constants. 
Let us assume that space-time is compact\footnote{If not, we 
restrict to the corresponding local consideration.}. We denote the 
volumes of 
$d\sigma^2$ by $V_1$ and of $d\tau^2$ by $V_2$, i.e.   
\be
V_1\quad = \quad  \int  \sqrt{- \det g_{AB}}\, d^2 x^A \, , \qquad 
 V_2\quad = \quad  \int  \sqrt{ \det g_{\alpha\beta}}\, d^2 x^{\alpha}
\ee
Due to the Gauss-Bonnet theorem we have two topological invariants
 which do not change by a smooth variation of the metric: 
\be
K_1\quad =
 \quad    \int  \   P  \   \sqrt{- \det g_{AB}}\, d^2 x^A \, , \qquad 
K_2\quad = 
\quad    \int  \   Q  \    \sqrt{ \det g_{\alpha\beta}}\, d^2 x^{\alpha}
\ee
Because of the constancy of $P$ and $Q$ we have 
$P=K_1/V_1$ and $Q = K_2/V_2$.  
We insert eqs. (11,13,14) into eq. (12) and get 
\be
I \quad = \quad K_1 \, V_2\ + \ K_2 \, V_1 \ - \ 2 \Lambda \, V_1 \, V_2
\ee
Extremality of the action $I$ implies $\frac{\partial I}{\partial V_n} =
0$ for
$n=1,2$: 
$$
0=K_2-2\Lambda V_2, \qquad 0=K_1-2\Lambda V_1 \, .$$ 
These equations imply $P=Q=2\Lambda$ and  
$R=4 \Lambda$.\footnote{In the usual deduction we get from the 
action (12) the
Einstein 
equation 
$$
R_{ij}- \frac{R}{2}g_{ij}= - \Lambda g_{ij} \, ,
$$
i.e., $R_{ij}=\Lambda g_{ij}$ and $R=4\Lambda$, whose spherically 
symmetric solution are almost everywhere given in Schwarzschild 
coordinates
as
$$
ds^2=- \left(1-\frac{2m}{r}-\frac{\Lambda}{3}r^2 \right)  dt^2  +  
\left(1-\frac{2m}{r}-\frac{\Lambda}{3}r^2 \right)^{-1} dr^2 + 
r^2 d\Omega^2
$$
 } 
For $\Lambda =0$ we get only the flat Minkowski space-time. 
For  $\Lambda>0$ however, we get a nonflat spherically symmetric 
space-time
\be
ds^2\quad = \quad \Lambda^{-1} \left[  - \, t^{-2}dt^2 + t^2dx^2   
+d\Omega^2
  \right] 
\ee
(where $d\Omega^2 = 
 d\theta^2 + \sin^2 \theta d \varphi^2 $ is the metric of the standard 
2-sphere) 
representing a cartesian product of two spaces of  equal positive 
constant curvature which is non-singular and not asymptotically flat. 
 Metric (16) represents a cosmological model of Kantowski-Sachs type
 and
possesses a 6-dimensional isometry group including a time-like 
isometry, 
(i.e.,
 the time-dependence of metric (16) is only due to the choice of the 
coordinates).

\medskip

Analogously we get for the case   $\Lambda<0$ the solution 
\be
ds^2\quad =
 \quad \vert \Lambda  \vert ^{-1} \left[  - \,x^{2}dt^2 +x^{-2}dx^2   
+d\bar\Omega^2
  \right] 
\ee
(where $d\bar\Omega^2 = 
 d\theta^2 + \sinh^2 \theta d \varphi^2 $ is the metric of the standard 
plane of constant negative curvature) representing a cosmological
 model
of Bianchi type III. 

\subsection{Curvature for this type of metrics}
For metric (7), the non-vanishing components $R_{ijkl }$ of the
 Riemann  tensor are
\be
R_{ABCD}\quad =
  \quad \frac{P}{2}(g _{AC}g_{BD}\ - \ g _{BC}g_{AD})
\ee
and
\be
 R_{\alpha\beta\gamma\delta}\quad = \quad \frac{Q}{2}(g _{\alpha
\gamma}g_{\beta \delta} \ - \ g _{\beta\gamma}g_{\alpha \delta} ) 
\ee
For the Ricci tensor we get analogously $R_{\alpha A}=0$, and
\be
R_{AB}\quad = \quad \frac{P}{2}g_{AB}\, , \qquad 
R_{\alpha\beta}\quad = \quad \frac{Q}{2}g_{\alpha\beta}
\ee
The Weyl tensor reads\footnote{The definition is
$$
C_{ijkl}=R_{ijkl}-\frac{1}{2}(R _{ik}g_{jl}+R _{jl}g_{ik}
-R_{jk}g_{il}-R_{il}g_{jk})+
\frac{R}{6}(g_{ik}g_{jl}-g_{jk}g_{il})
$$
}  
\be
C_{\alpha\beta\gamma\delta}\quad = \quad 
\frac{R}{6}(g_{\alpha\gamma}g_{\beta\delta} \ - \ g_{\beta\gamma}
g_{\alpha\delta})
 \ee
\be
C_{ABCD}\quad = \quad 
\frac{R}{6}(g_{AC}g_{BD} \ - \ g_{BC}g_{AD})
 \ee
As a byproduct we get:   conformal flatness of metric (7) implies 
the vanishing of its curvature scalar $R$. 
 However, in contrast to the Riemann tensor, the Weyl tensor possesses 
also
 non-vanishing mixed components: 
\be
C_{\alpha A \beta B} \quad=\quad - \frac{R}{12}g_{\alpha\beta}g_{AB}
\ee
Summing up eqs. (21-23) we get
$$
C_{ijkl} C^{ijkl} \ = \   \frac{1}{3} \, R^2
$$

\subsection{Solving the Bach equation  -- constant $P$ and $Q$  }
As first part we make the analogous calculation as in section 4.2.
Inserting eqs.(11,18-20) into eq. (3) we get $L_{GB} = 2PQ$, i.e., the
4-dimensional topological
invariant is the double product of the corresponding 2-dimensional ones:
\be
 \int L_{GB}
 \sqrt{- \det g_{ij}}\, d^4 x^i \quad=\quad 2 \ K_1 \ K_2
\ee
Further we get $R^2=(P+Q)^2$ and
\be
R_{ij} \ R^{ij} \quad=\quad \frac{1}{2} \ (P^2 \ + \ Q^2)
\ee
thus, up to a divergence, eq. (1) now reads 
\be
\hat L 
\quad=\quad  \frac{1}{3} \ (P^2 \ + \ Q^2)
\ee
and with the notation from eq. (13) and 
\be
L_1 \  = 
 \    \int  \   P^2  \   \sqrt{- \det g_{AB}}\, d^2 x^A \, , \qquad 
L_2 \ 
 =  \  \int  \  Q^2  \  \sqrt{ \det g_{\alpha\beta}}\, d^2 x^{\alpha}
\ee
\be
\hat I \equiv   \int \hat L 
 \sqrt{- \det g_{ij}}\, d^4 x^i=\frac{1}{3} \ (L_1V_2 +L_2 V_1)
\ee
In the set of spaces with constant $P$ and $Q$ we get for $n=1,2$: 
 $ \, L_n =K_n^2/V_n$, i.e. 
\be
\hat I \quad=\quad \frac{1}{3} \ \left(K_1^2 V_2/V_1 \ 
 + \  K_2^2 V_1/V_2 \right)
\ee
Consequently, 
$$
\frac{\partial \hat I}{\partial V_n} =0
$$
implies $0=K_1^2/V_1 - K_2^2V_1/V_2^2$, i.e. 
 $P^2 \  = \  Q^2$. Thus, besides the Einstein spaces of this type we 
additionally 
get spaces with $P = - Q \ne 0$. These are just 
 the cartesian products of two 2-spaces of constant 
non-vanishing curvature with the additional condition that they 
have $R=0$, i.e., that 
they are conformally flat, which can be shown by eqs. (21-23) . 

By use of the notation of sct. 3 we can say that 
 they represent trivial solutions of the Bach equation.   

\subsection{Solving the Bach equation  -- variable $P$ or $Q$  }
From eq. (25) we see: $R_{ij}R^{ij} - \frac{1}{2}R^2$ represents a 
divergence.\footnote{In the 2-parameter class of Lagrangians
  $L_{\alpha,\beta}= \alpha R_{ij}R^{ij} + \beta R^2$, the 
case $\alpha + 3 \beta=0$
 leads to Weyl gravity, cf. sct. 2;  and the Eddington case is defined 
by $\alpha + 2 \beta=0$, 
 this case we meet here,  cf. [10] for details.} Thus, it seems tempting 
 to use also this divergence to show that our Lagangian gives just the 
same 
 field equation than $L=R^2 $ would  give. But this argument does 
not work
 from the following reason:
 The statement, that  $R_{ij}R^{ij} - \frac{1}{2}R^2$ 
represents a divergence, 
 is valid  only within the class of metrics
 considered here, so   the vacuum field equation need not be  the same
 for $L$  eq. (1) and the Lagrangian $R^2$: The variation has to be made 
 in comparison with all possible metrics. Therefore, we have now to 
use the
full equation (2) or (4-6). For our purposes it turned out that eq. (2) is
easier to handle.   
 We  write 
 \be
\Box R \quad=\quad \Box P \ + \ \Box Q  
\ee
with a context-dependent meaning of the symbol $\Box$, cf. eqs. (7-11). 
E.g.: In $\Box P$, $\Box$ denotes the D'Alembertian within $d\sigma^2$.
 Analogously, we use only one symbol ``;" for the covariant derivative. 
 After a lenghty but straightforward calculation we get 
 the $AB$-part of the Bach equation:  
\be
B_{AB} \  \equiv
  \  \frac{1}{3}P_{;AB} + \ g_{AB}
\left(\frac{1}{6}\Box Q - \frac{1}{3}
 \Box P + \frac{1}{12} Q^2 - \frac{1}{12}P^2
\right) \ = \ 0 
\ee
From the trace of  eq. (31) we see that  $\Box P + \frac{1}{2} P^2$
 and  $\Box Q+ \frac{1}{2} Q^2$
 are both constant because
they are equal but ``live'' in different spaces:
 \be
\Box P \, + \,  \frac{1}{2} P^2\quad=\quad   C,
 \qquad \Box Q\, + \,  \frac{1}{2} Q^2 \quad=\quad  C \quad=\quad const.
\ee
The fact that the trace-free part of the tensor $P_{;AB}$ vanishes is 
equivalent 
to the requirement that $\xi^A =
\varepsilon^{AB}P_{;B}$ represents a Killing vector.\footnote{We
use $\varepsilon^{AB}$, the covariantly constant  antisymmetric
Levi-Civita pseudotensor in $d\sigma^2$.}

Now we have to use the $ \alpha \beta$-part of the field equation. 
However, 
we need not really deduce it, because there is a duality 
$A \leftrightarrow \alpha$. Thus, the only additional requirement is that 
$\eta^{\alpha} = \varepsilon^{\alpha \beta} Q_{; \beta}$ represents a
Killing vector, too. 

\bigskip

Let us summarize this section: 
1. There exists a double-Birkhoff theorem as 
follows: 
If a solution of the Bach equation is the cartesian product of two 
2-spaces, 
then
 2 independent Killing vectors exist. 
 They are orthogonal to each other, and each of them is
 hypersurface 
orthogonal.
If either $P$ or $Q$ is constant, then the number of Killing vectors 
equals 4.

\medskip

2. The cartesian product of two 2-spaces is a solution of the Bach equation 
if and
only if there exists a constant $C$ such that both 2-spaces solve the 
fourth-order
field equation following from the 2-dimensional 
Lagrangian\footnote{Here, ${}^{(2)}R$  is the curvature scalar in 
that 2-dimensional space.
 In principle, this result could have been guessed already from eq. (26): 
In the variation of
(26) with respect to $g_{AB}$, the scalar $Q^2$ plays the role of a 
constant and vice versa. However, by such a consideration one looses
 the information about 
the fact, that both equations (32) contain the {\it same} constant $C$.}
   $L \ = \  \frac{1}{2} \  {}^{(2)}R^2 \,  + \, C$.

\medskip

3. The solutions for   
 $L \ = \  \frac{1}{2} \ {}^{(2)}R^2 \,  + \, C$ are all known in closed 
form
 [8, eq.(14)],\footnote{With the ansatz
 $d{}^{(2)}s^2=dw^2/A(w) \pm A(w)dy^2$
 one gets -- besides the constant curvature 
 solution ${}^{(2)}R^2 \equiv 2C $   --
the general solution as $A(w)=C_1 +Cw - w^3/6$ where $C_1$
is a further constants. One should note that the cubic term in $A(w)$  
 is necessarily unequal zero 
to have a non-constant curvature scalar, and that therefore, a term 
$\approx w^2$ can be made vanish by a suitable $w$-translation.
 However, the fact that the cubic term is just $-1/6$ was fixed by a 
corresponding 
coordinate transformation, a multiplication of $w$ and $y$ by the same 
constant.  This ``$-1/6$''  was chosen such that the factor $C$ of the 
linear term
 is just the $C$ defined in eq. (32).}
 so we are now able to list all these solutions of the Bach equation
 which possess exactly 2 Killing vectors:
 \be
-(a+Cr-r^3/6)dt^2 + \frac{dr^2}{a+Cr-r^3/6}+
(b+C\psi- \psi^3/6) d\phi^2+ \frac{d\psi^2}{b+C\psi- \psi^3/6}
\ee
with 3 constants $ a, \, b, \, C$.  Each of the two factor spaces 
gives one  integration constant, but from eq. (32) it follows, that
  the  third constant reflects that fact that the Bach equation is 
scale-invariant. 

\medskip

4. The number of Killing vectors of a solution of the Bach equation 
 for the metrics discussed here equals 2, 4, 6, or 10.  
The solutions with 6 Killing vectors and the flat space-time solution  with 
10 Killing vectors have already been listed in sct. 4.4, the solutions with
2 Killing vectors are given by eq. (33) above; thus,  it remains to find the
solutions possessing 4 Killing vectors. This takes place if from $P$ and $Q$ 
one is constant, 
and the other one not. For $C<0$, all solutions have exactly 2 Killing 
vectors because 
neither $P$ nor  $Q$ can be const., cf.  eq. (32).
 We restrict to the case that $Q$ is constant, and $P$ not constant; the 
other case is quite analogously to deal. Depending on the sign of $Q$, 
we have 
3 different subcases. 

\medskip

 The spherically
 symmetric solutions we get  for the case that $d\tau^2=d\Omega^2$,
i.e., $Q=C=2$.\footnote{With $Q=-2,
 \ C=2$ and $d\tau^2=d\bar\Omega^2$ we get
the analogous case for a plane of negative curvature. The formulas
 can 
be 
straightforwardly
 written down.}   
 \be
ds^2 =-(a+2r - r^3/6)dt^2 + \frac{dr^2}{a+2r -r^3/6}+d\Omega^2
\ee
This is -- up to conformal transformations --
 the general spherically symmetric solution of the Bach equation, 
however,
this form of the solution is not very common. Therefore, 
we multiply metric (34)
by a conformal factor $\rho^2(r)$ and use $\rho = c \pm 1/r$ as new 
radial coordinate. 
As a result one gets the known old result (see [6] also for the
details of that transformation) that the spherically 
symmetric solutions of the Bach equation are conformally related to the
Schwarzschild--de Sitter solution.  

\medskip

 The plane-symmetric
 solutions we get for $d\tau^2=dy^2+dz^2$, i.e. $Q=C=0$.
 \be
ds^2 \  = \ -(a- r^3/6)dt^2 + \frac{dr^2}{a - r^3/6}+
 dy^2+ dz^2
\ee

\section{Cosmological solutions}

Here we give some examples of cosmological solutions of the type 
eq. (34) and
 (35). The interpretation of the more general solution (33) as 
cosmological model 
 shall be postponed to later work.

\subsection{Axially symmetric Bianchi type I Universe}

In order that to obtain some cosmological solutions in conformal 
Weyl gravity we have to do the following. The 2-metric 
$d\sigma ^2$ for metric (35)  can be written as: 
\begin{equation}
d\sigma ^2 = -\left( a + br^3 \right)dt^2 + 
\frac{dr^2}{a + br^3} .
\label{36}
\end{equation}
 (Because of $C=0$ the factor $b$ need not be put to $-1/6$.) 
It is evident  that for $a,b<0$ we can rename 
$t \rightarrow x$, $r \rightarrow t, a\rightarrow -a, b\rightarrow -b$
 then we have:
\begin{equation}
d\sigma ^2 = -\frac{dt^2}{a + bt^3} + \left( a + bt^3 \right)dx^2 .
\label{37}
\end{equation}
Let we introduce the polar coordinate system 
$y = \rho\cos \varphi$, $z = \rho\sin\varphi$ then the  solution (35)
is given by: 
\begin{equation}
ds^2 = -\frac{dt^2}{a + bt^3} + \left( a + bt^3 \right)dx^2 + 
d\rho ^2 + \rho ^2 d\varphi ^2 .
\label{38}
\end{equation}
The metric 
(\ref{38}) describes the axially  symmetric Kasner-like Universe 
with expanding $x$-dimension and constant $(y,z)-$plane. 
\par 
The calculations of the scalar invariants for this metric 
give us: 
\begin{eqnarray}
R & = & -6 b t,
\label{39}\\
R_{ik}R^{ik} & = & 18b t^2 ,
\label{40}\\
R_{iklm}R^{iklm} & = & 36 b t ^2 .
\label{41}
\end{eqnarray}
At the surface defined by  $t= - (a/b)^{1/3}$ there is only a coordinate 
pecularity similar to that one of the Schwarzschild horizon.  

\subsection{Spherically symmetric Universe}

Analogously to the previous case we can exchange 
$t \rightarrow r$ and $r \rightarrow t$ in the solution (34)  and then
 we get: 
\begin{equation}
ds ^2 = - \frac{dt^2}{a + 2t - t^3/6} + 
\left( a + 2t - t^3/6 \right) dr^2 + d\Omega ^2 .
\label{42}
\end{equation}
The scalar invariants are: 
\begin{eqnarray}
R & = & t,
\label{43}\\
R_{ik}R^{ik} & = & \frac{1}{2} t^2 ,
\label{44}\\
R_{iklm}R^{iklm} & = & t^2 .
\label{45}
\end{eqnarray}
Also, at  $t_0$ (where $a + 2t_0 - t^3_0/6 = 0$) there is not 
a real singularity and we have only the peculariaty 
$g_{rr} = 0$ and $g_{tt} = - \infty$.

\section{Discussion}
In many papers,  motivations for considering  conformal  Weyl
 gravity, i.e., motivations for solving the Bach equation, are given. 

\medskip

The Bach tensor (sometimes also called: 
Schouten-Haantjes tensor) plays
 also a role in the following contexts:

\medskip

1. The integrability of the null-surface formulation of General 
Relativity
 imposes
 a field equation on the local null surfaces which is equivalent 
to the 
vanishing 
of the Bach tensor, see [11]. 

\medskip

2. For asymptotically flat space-times it holds: It is conformally
related to a Ricci-flat space-time if and only if the Bach tensor 
vanishes, see [12] .

\medskip

3. The Mannheim-Kazanas approach [4], see also  its analysis in [13], 
essentially
uses the Bach tensor and tries to relate it to observable astrophysical 
effects. 

\medskip 

4. The Bach equation accompanied by conformally invariant matter 
(electromagnetic field) has been discussed in [14]. There, already the 
relation 
  to the $R^2$-gravity in 2 dimensions has been mentioned, and a 
Birkhoff theorem has been deduced. However, at that time, the
 solutions 
 of  $R^2$-gravity in 2 dimensions (deduced in [15]) were not known, 
so this relation 
was not so useful as it is now. 

\medskip

5. In several approaches to quantum gravity, e.g. by compactification 
of 
11-dimensional supergravity [16], one gets $R^2$-terms including 
the 
Weyl-term in the effective action. In [17], a theorem relating solutions 
of a 
4-order theory of gravity to General Relativity has been deduced. 
 In both the papers [16, 17], the general need to include the Weyl term
 is mentioned, 
but the calculations have been restricted  to the case where this term is 
absent.  
 So, here remains a general interest in these calculations, too. 

\medskip

6. Quite recently, renewed interest in  the Bach equation
 lead to several concrete calculations, see [18], [19]  and the 
references 
cited there. 
 In [18], a new approach to the Newtonian limit of conformal gravity 
has been 
presented,   and in [19], the   Hamiltonian formulation
and exact solutions of the Bianchi type I spacetime in conformal
gravity have been deduced. The exact solutions given there are more 
general 
 for the Bianchi type I case than our cosmological solution given 
in sct. 5.1.
 However,  our general solution   (33)  possessing only 2 Killing
 vectors is  
more 
general than the solutions given in  [19].

\section*{Acknowledgement}
Financial support from DFG and A.-v.-Humboldt-Found. 
is gratefully acknowledged.
We thank the colleagues of Free University Berlin, 
 especially Prof.  H. Kleinert, for valuable comments.

\section*{References}

\noindent
[1] R. Bach, Math. Zeitschr. {\bf 9} (1921) 110.

\medskip

\noindent
[2] H. Weyl, Sitzber. Preuss. Akad. d. Wiss. Berlin,
 Phys.-Math. Kl. (1918) 465. 

\medskip

\noindent
[3] H.-J. Schmidt, Astron. Nachr. {\bf 307} (1986) 339.

\medskip

\noindent
[4] P. Mannheim, D. Kazanas, Gen. Relat. Grav. {\bf 26}
 (1994) 337. 

\medskip

\noindent 
[5] C. Kozameh, E. Newman, K. Tod,  Gen. Relat. Grav. {\bf 17}
 (1985) 343. 

\medskip

\noindent 
[6] H.-J. Schmidt:  A new conformal duality of spherically symmetric 
space-times,  gr-qc/9905103 and references cited there. 

\medskip

\noindent
[7] L. Querella:  Variational principles and 
cosmological models in higher-order gravity, Doct. Diss.
Liege 1998, 194 pages, gr-qc/9902044;  R. Schimming, p. 39 in: 
M. Rainer, 
H.-J. Schmidt (Eds.)
Current topics in mathematical cosmology, WSPC Singapore 1998.

\medskip

\noindent
[8] H.-J. Schmidt: The classical solutions of two-dimensional
gravity, 1999, Gen. Relat. Grav. {\bf 31} 1187, cf.  gr-qc/9905051, 

\noindent 
http://www.physik.fu-berlin.de/\~{}hjschmi/tex/s107.tex \quad, 

\noindent  
 http://www.physik.fu-berlin.de/\~{}hjschmi/tex/s107.ps

\medskip

\noindent 
[9]  A. Krasinski, Gen. Relat. Grav. {\bf 31} (1999), 945.

\medskip

\noindent 
[10] H.-J. Schmidt, Ann. Phys. (Leipz.) {\bf  44} (1987) 361.

\medskip

\noindent 
[11] M. Iriondo, C. Kozameh, A. Rojas: Null surfaces 
and the Bach equation,
 J. Math. Phys. {\bf 38} (1997) 4714.

\medskip

\noindent 
[12]
L. Mason, J. Math. Phys. {\bf 36} (1995) 3704. 

\medskip

\noindent 
[13] V. Perlick, C. Xu, Astrophys. J. {\bf 449} (1995) 47. 

\medskip

\noindent 
[14] R. Riegert, Phys. Rev. Lett. {\bf 53} (1984) 315. 

\medskip

\noindent 
[15] H.-J. Schmidt, J.
Math. Phys. {\bf 32} (1991) 1562.

\medskip

\noindent 
[16] J. Ellis, N. Kaloper, K. Olive, J. Yokoyama, Phys. Rev. D
 {\bf 59} (1999) 103503. 

\medskip

\noindent 
[17] D. Barraco, V. Hamity, 
Gen. Relat. Grav. 31 (1999) 213.

\medskip

\noindent 
[18] O. Barabash, Yu. Shtanov,
 Phys. Rev. D in print.

\medskip

\noindent 
[19]
J. Demaret, L. Querella, C. Scheen, Class. Quant. Grav. 16 (1999) 749. 

\end{document}